\documentclass[10pt, conference, letterpaper]{IEEEtran}

\makeatletter
\def\ps@headings{%
\def\@oddhead{\mbox{}\scriptsize\rightmark \hfil \thepage}%
\def\@evenhead{\scriptsize\thepage \hfil \leftmark\mbox{}}%
\def\@oddfoot{}%
\def\@evenfoot{}}
\makeatother 

\usepackage{mathrsfs}
\usepackage{amsfonts}
\usepackage{array}
\usepackage{amssymb}
\usepackage{amsmath}
\usepackage{graphicx}
\usepackage{multirow}
\usepackage{amsthm}

\usepackage[lined,vlined,ruled,commentsnumbered]{algorithm2e}
\usepackage{epstopdf}
\usepackage{stmaryrd}
\usepackage{multirow,url}
\usepackage{color,cite}
\usepackage{slashbox}
\usepackage{bm}
\usepackage{indentfirst}
\usepackage{epsfig}

\usepackage[table]{xcolor}

\usepackage{fancyhdr}

\fancypagestyle{firststyle}
{
   \fancyfoot[C]{\rev{xxx-x-xxxx-xxxx-x/xx/\$31.00 \copyright 2015 IEEE}}
}

\IEEEoverridecommandlockouts

\newtheorem{theorem}{Theorem}
\newtheorem{lemma}{Lemma}

\theoremstyle{plain}

\ifodd 1212
\newcommand{\rev}[1]{{\color{blue}#1}}

\newcommand{\comg}[1]{\textbf{\color{green} (COMMENT: #1)}}
\newcommand{\response}[1]{\textbf{\color{green} (RESPONSE: #1)}}
\else
\newcommand{\rev}[1]{{#1}}

\newcommand{\comg}[1]{}
\newcommand{\response}[1]{}
\fi

\newcommand{\revv}[1]{{#1}}
\newcommand{\revvv}[1]{{#1}}

\def\eq{\triangleq}
\def\dd{\mathrm{d}}
\def\txsty{ }

\def\sumN{\sum_{n\in\N} }
\def\sumI{\sum_{i\in\I} }
\def\sumT{\sum_{t\in\T} }
\def\intTH{\int_{\th\in\TH}}

\def\N{\mathcal{N}}
\def\I{\mathcal{I}}
\def\T{\mathcal{T}}
\def\AR{\mathcal{A}}
\def\ARi{\mathcal{A}_i}

\def\w{w}
\def\wi{\w_i}
\def\wit{\w_i[t]}

\def\y{y}
\def\yi{\y_i}
\def\yit{\y_i[t]}

\def\v{v}

\def\vnt{\v_n[t]}
\def\c{c}
\def\cn{\c_n}
\def\cnt{\c_n[t]}

\def\C{C} 	
\def\V{V} 	
\def\Ct{\C[t]} 	
\def\Vt{\V[t]} 	
\def\VC{S} 			
\def\VCt{\VC[t]} 	

\def\bid{\c^{\prime}}
\def\bidnt{\bid_n[t]}
\def\regu{\mu}
\def\bregu{\boldsymbol{\regu}}
\def\regun{\regu_n}
\def\pnt{p_n[t]}

\def\VCo{\VC^{\circ}} 			
\def\VCso{\VC^{*}} 			
\def\VClo{\VC^{\dag}} 			

\def\wc{\widetilde{\c}} 	
\def\wcnt{\wc_n[t]} 	
\def\WVC{\widetilde{\VC}} 	

\def\x{x}	
\def\xn{\x_n}	
\def\xnt{\x_n[t]}
\def\bx{\boldsymbol{\x}}
\def\bxt{\bx[t]}
\def\bxn{\bx_n}
\def\bX{\mathbf{\x}}

\def\xnlo{\x^{\dag}_n}
\def\bxlo{\bx^{\dag}}
\def\bxnlo{\bxn^{\dag}}

\def\xnau{\x^{\ddag}_n}
\def\bxau{\bx^{\ddag}}

\def\z{z}
\def\zni{\z_{n,i}}
\def\znit{\z_{n,i}[t]}
\def\bz{\mathbf{\z}}

\def\bznt{\bz_{n}[t]}

\def\d{d}	
\def\dn{\d_n}	
\def\D{D} 	

\def\dnlo{\dn^{\dag}}	

\def\th{\theta}
\def\tht{\th[t]}

\def\TH{\Theta}

\def\que{q}
\def\quen{\que_n}
\def\bque{\boldsymbol{\que}}
\def\lyp{J}
\def\drift{\Delta}
\def\pnlt{\Pi}
\def\lypV{\phi}

\begin{document}

\title{Providing Long-Term Participation Incentive in Participatory Sensing
\vspace{-8mm}
}


\author{Lin~Gao,
        Fen~Hou,
        and~Jianwei~Huang
\thanks{\rev{This work is supported by ...}}
\thanks{L. Gao and J. Huang are with Dept.~of Information Engineering, The Chinese University of Hong Kong, HK,
Email: \{lgao, jwhuang\}@ie.cuhk.edu.hk;
F. Hou is with Dept.~of Electrical and Computer Engineering, University of Macau, Macau. Email: fenhou@umac.mo}
}

\addtolength{\abovedisplayskip}{-1mm}
\addtolength{\belowdisplayskip}{-1mm}

\maketitle

\begin{abstract}
Providing an adequate long-term participation incentive is important for a participatory sensing system to maintain enough number of active users (sensors), so as to collect a sufficient number of data samples and support a desired~level~of service quality.
In this work, we consider the sensor selection problem in a general time-dependent and location-aware participatory sensing system, taking     the long-term user participation incentive into explicit consideration.
We study the problem systematically under different information scenarios,  regarding both future information and current information (realization).
In particular, we propose a Lyapunov-based VCG auction policy for the on-line sensor selection, which converges asymptotically to the optimal off-line benchmark performance, even with no future information and under (current) information asymmetry.
Extensive numerical results show that our proposed policy outperforms the state-of-art policies in the literature, in terms of both user participation (e.g., reducing the user dropping probability by $25\%\sim90\%$) and social performance (e.g., increasing the social welfare by $15\%\sim80\%$).~~~~

\end{abstract}

\IEEEpeerreviewmaketitle

\thispagestyle{firststyle}


\section{Introduction}\label{sec:introduction}

\subsection{Background and Motivations}


The proliferation of mobile devices (e.g., smartphones) with rich embedded sensors has led to revolutionary new sensing paradigm, often known as \emph{Participatory Sensing} \cite{phone-survey,Burke-ps, Ganti-crowdsensing}, in which mobile users voluntarily participate in and actively contribute to sensing system, using their carrying smartphones. 
Due to the low deploying cost and high sensing coverage, this new paradigm has attracted a wide range of applications in environment,  infrastructure, and community  monitoring (e.g., 
air pollution \cite{App-OpenSense, App-Atmos1, App-WeatherLah}, 
wireless signal strengths \cite{App-OpenSignal, App-NoiseTube, App-Sensorly}, 
road traffic \cite{App-Waze, App-Millennium, App-CarTel}, and parking \cite{App-SpotSwitch, App-ParkNet}).

A typical participatory sensing system architecture usually consists of a service platform (also called \emph{service provider}) residing in the cloud and a collection of mobile smartphone users 
\cite{yang-mobicom12, zhuym-infocom14, zhuym-infocom14-b}. 
The service provider launches a set of sensing tasks with different sensing requirements for different purposes, and mobile users actively subscribe to (participate in) one or multiple sensing task(s). 
In this work, we focus on an important type of participatory sensing scheme called the \emph{server-initiated} sensing, where the service provider selects a specific set of participating smartphones to perform the sensing task, depending on the spatio-temporal data requirement of the sensing task and the geographical locations of the participating users as well as their sensing capabilities. 
Comparing with the \emph{user-initiated} sensing scheme (where users actively decide when and where to sense), the server-initiated sensing scheme gives more control to the service provider to decide when and where to collect the data at what costs, hence can better fit the requirements of sensing tasks. 
Clearly, the success of such a sensing system strongly relies on the active participations of users as well as their willingnesses to contribute their sensing capability and resource to the sensing tasks. 

Although many participatory sensing applications 
have been proposed in \cite{App-WeatherLah, App-OpenSense, App-Atmos1, App-OpenSignal, App-NoiseTube, App-Sensorly, App-Waze, App-Millennium, App-CarTel, App-SpotSwitch, App-ParkNet}, 
they simply assume that users \emph{voluntarily} participate in the system to perform sensing tasks. 
In reality, however, users may not be willing to participate in the sensing system, as this will incur extra operational cost (e.g., the battery energy expenditure and the transmission expense). 
Moreover, many sensing tasks are location-aware and time-dependent, and involve spatio-temporal context. 
Sharing sensing data tagged with spatio-temporal context may reveal a lot of personal and sensitive information,~which poses potential \emph{privacy} threats to the participating users \cite{Privacy-Cristo2013}.
All of these bring the \emph{incentive} issue to the fore. 

Several recent works have been devoted to the incentive mechanism design issue in participatory sensing, mainly using pricing 
and auction \cite{yang-mobicom12, Lee-2010, Lee-2010-b, zhuym-book2013, Jaimes-2012, Koutsopoulos-infocom13, zhuym-infocom14-b, Luo-infocom2014, Duan-infocom2012}.
Most of them focus on compensating the user's \emph{direct} sensing cost when being chosen as a sensor to perform a particular sensing task (e.g., in \cite{yang-mobicom12, zhuym-book2013, Jaimes-2012, Koutsopoulos-infocom13, zhuym-infocom14-b, Luo-infocom2014, Duan-infocom2012}), 
which we call the \emph{short-term sensing incentive}. 
In practice, however, we find that the users participating in a sensing task may suffer certain \emph{indirect} cost even when not performing the sensing task.\footnote{For example, in a location-aware sensing task, users need to periodically report their locations to the service provider before the latter makes the sensor selection decision, which incurs certain energy and transmission cost.
} 
In this case, the short-term sensing incentive may \emph{not} be enough to guarantee the long-term continuous participations of users. 
Intuitively, if a user is rarely selected as a sensor (hence hardly receives the short-term sensing incentive), the user may lose the interest in continuous participation and decide to drop out of the sensing system (e.g., shut down the sensing app on his smartphone). 
Without an adequate number of users participating in the system, however, the service provider may not be able to collect a sufficient number of sensing data to support a desired service quality (e.g., miss the road traffic informaiton in some areas).~~~~~~~~~~~~~~~

To the best of our knowledge, \cite{Lee-2010} and \cite{Lee-2010-b} are the~only results that explicitly study the \emph{long-term participation incentive}  in participatory sensing. 
To stimulate the continuous participation of users, Lee \emph{et al.} in \cite{Lee-2010} and \cite{Lee-2010-b} introduce a \emph{virtual credit} for lowering the bids of users who lost in the previous round of auction, hence increasing their winning probabilities in future auction rounds. 
However, they consider neither the truthfulness, nor the optimality of the proposed auction. 
In this work, we will study the {long-term participatory incentive}, joint with the {short-term sensing incentive}, with rigorous truthfulness and optimality analysis.



\begin{figure}[t]
\centering
\includegraphics[scale=0.4]{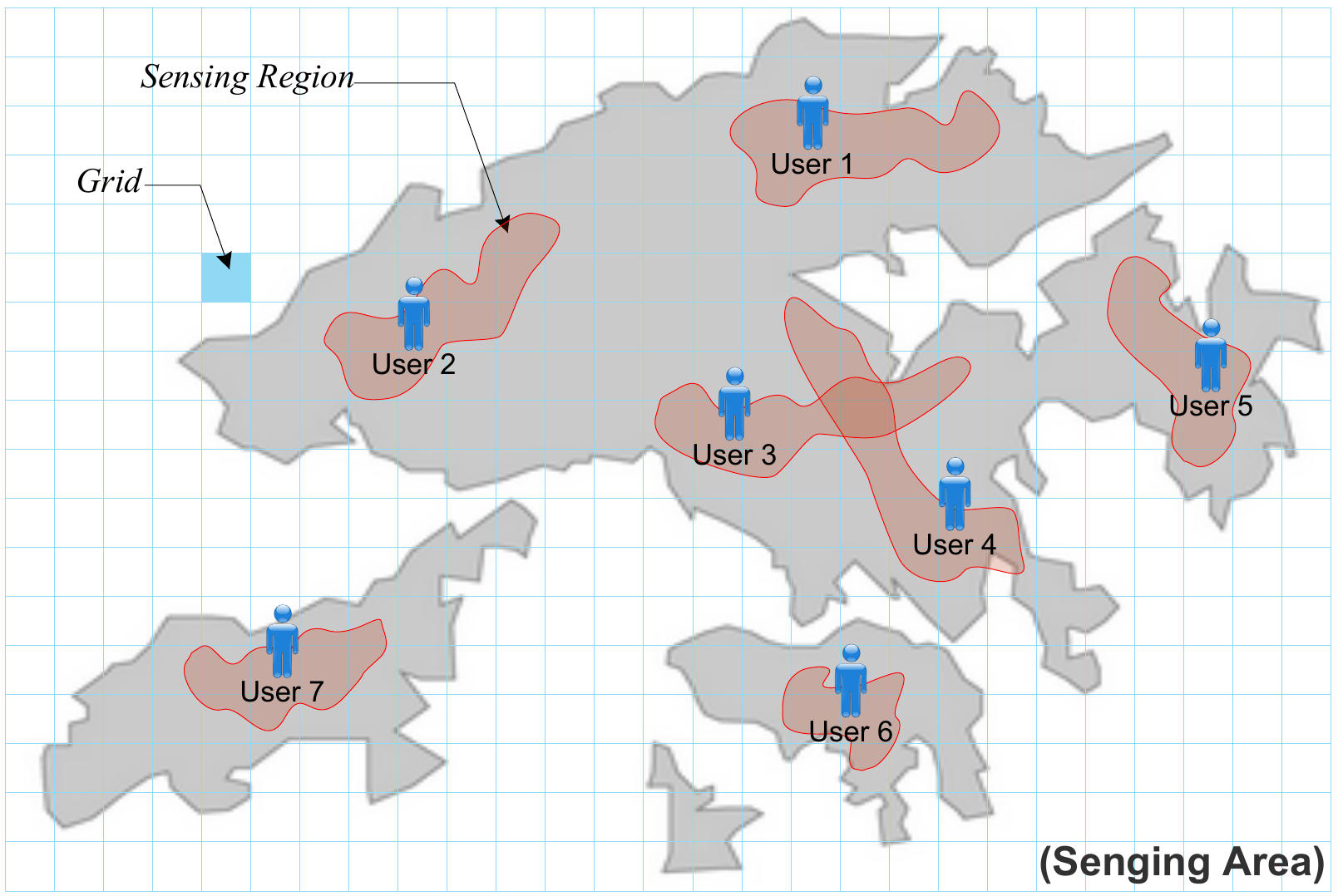}
\caption{\rev{Location-Aware Participatory Sensing Model.} 
} \label{fig:model}
\vspace{-3mm}
\end{figure}

\subsection{Solutions and Contributions}

Specifically, we consider a general location-aware, time-dependent participatory sensing system, where the data in different time (slots) and/or locations may have different values for the sensing tasks.
\rev{Each participating user has the potential to sense a specific region (at a certain sensing cost) in a specific time slot, depending on his location and mobility pattern.} 
Fig.~\ref{fig:model} illustrates a snapshot of such a sensing   system (in a particular time slot), where the sensing region of each user is denoted by the shadow area around the user. 
%
In such a system, 
the service provider selects (allocates) users as sensors to perform sensing tasks slot by slot. 
We focus on the following \emph{sensor selection$/$allocation problem} for the service provider:

\noindent	
$~~~~\bullet$
\emph{Which users should be selected as sensors in each time slot, aiming at maximizing the social welfare and ensuring the long-term participation incentive of users?}

\noindent
The problem is challenging due to the following reasons. 
\rev{First, the overlap of different users' sensing regions makes  their sensing activities possibly redundant (hence partially ``conflict'' with each other). 
Second, the long-term participation incentive of users makes the sensor allocations in different time slots coupled.} 
Based on the above, our model and problem formulation capture the following important features of a participatory sensing system: 
(i) long-term participation incentive, (ii) time-dependent and location-aware sensing requirement, and (iii) partial conflicting sensing activity. 
As far as we know, this is the first work that systematically studies a participatory sensing problem with all of the above features.

\revv{We solve the above sensor selection problem under different information scenarios, regarding both \emph{future} information (i.e., complete, stochastic, or no future information) and \emph{current} information (i.e., symmetric or asymmetric). 
Specifically, with complete or stochastic future information,  we formulate and solve an \emph{off-line} sensor selection problem as benchmark (where we assume that the current information is always symmetric). 
With no future information, we formulate and solve an \emph{on-line} sensor selection problem: (i) under information symmetry, we propose a \emph{Lyapunov-based} on-line sensor selection policy (Policy 1), which converges to the optimal off-line benchmark asymptotically;  
and 
(ii) under information asymmetry, we propose a Lyapunov-based \emph{VCG auction} policy (Policy 2), which is truthful, and meanwhile achieves the same asymptotically optimal performance as in Policy 1.}\footnote{\rev{Several recent works also studied the on-line policy for sensing task allocation, considering the uncertainty of user arrival \cite{infocom-reviewer-1,infocom-reviewer-2}. 
However, these works did not consider the user long-term participation incentive.}}

\rev{It is important to note that the key contributions of this work  are not on the Lyapunov framework itself, but rather, on the novel problem formulation and solution techniques.} 
For more clarity, we list the main results in Table~\ref{table:result}, and summarize the key contributions  as follows.
 

\noindent	
$~~~~\bullet$
\emph{Novel Model and Problem Formulation:} 
We study a general time-dependent  and location-aware participatory sensing system, taking into consideration the important but less studied issue of  long-term user participation incentive. 
We propose a simple yet representative formulation based on the allocation probability of each user to capture such an incentive. 

\noindent	
$~~~~\bullet$
\emph{Multiple Information Scenarios:} 
We study the optimal sensor selection problem under different information scenarios. 
In particular, we propose on-line sensor selection policies that converge to the asymptotically optimal performance, even with no future information and under information asymmetry.~~~~~~~~

\noindent	
$~~~~\bullet$ 
\emph{Performance Evaluations:} 
We compare the proposed on-line policies with the state-of-art policies, and show our proposed policies outperform the existing ones significantly, in terms of both  {user participation} and {social performance}: 
(i) Comparing with the RADP-VPC policy proposed in \cite{Lee-2010}\cite{Lee-2010-b}, our policies can reduce the user dropping probability by $25\%\sim 50\%$, and increase the social welfare by $15\% \sim 40\%$; 
(ii) Comparing with the Greedy/Random policy widely used in existing systems (e.g., \cite{App-Waze}), our policies can reduce the user dropping probability by $70\% \sim 90\%$, and increase the social welfare by $65\% \sim 80\%$. 


%

\begin{table}
\caption{Main Results in This Paper}
\label{table:result}
\hspace{-5mm}
\begin{tabular}{|m{1.5cm}|c|c|m{1.9cm}|c|}
\hline
\centering \textbf{Future Info.}  & \textbf{Current Info.}  & \textbf{Solution} & \centering \textbf{Performance} & \textbf{Section}
\\
\hline
\hline
\centering  Complete / Stochastic & Symmetric & Off-line & \centering  Optimal {\scriptsize(Benchmark)} & \ref{sec:solution}
\\
\hline
\centering No Info  & Symmetric  & On-line Policy 1 & \centering Asymptotic Opt. & \ref{sec:noinform}
\\
\hline
\centering No  Info  & Asymmetric  & On-line Policy 2 & \centering Asymptotic Opt. & \ref{sec:asymmetric}
\\
\hline
\end{tabular}
\vspace{-3mm}
\end{table}


\section{System Model}\label{sec:model}

We consider a location-aware participatory sensing system with a 
service provider (SP) and a set $\N \eq \{1,...,N\}$ of mobile smartphone users (participating in the system). 
The SP wants to collect specific data in a certain area (via participating users' smartphones) for specific tasks. 
\rev{Mobile users move randomly in and out of the desirable sensing area according to certain mobility pattern. 
As shown in Fig.~\ref{fig:model}, each user has the potential to sense a specific region in a certain period according to his location and mobility,}  
and the whole sensing area $\AR$ is divided into a set $\I =\{1, ... , I\}$ of grids.\footnote{A grid is the minimum unit of sensing area at a particular location, e.g., a square of $100\mathrm{m}\times 100 \mathrm{m}$, associated with a single data in a particular time.} 
Each grid $\ARi, i\in\I$, 
is associated with a weight $\wit$, denoting the \emph{value} of the data associated with grid $\ARi$ in each slot $t$.
Obviously, such a data value is location-aware and time-dependent. 

The SP requests data slot by slot, where each \emph{time slot} ranges from minutes to hours depending on the temporal data requirements of tasks. 
We consider the sensing operation in a long period consisting of a set $\T \eq \{1, ..., T\}$ of $T$ slots. 
At the beginning of each time slot, the SP selects (allocates) a set of users to perform the sensing task in that time slot, depending on factors such as the user locations and the data values.  
Let $\xnt\in \{0,1\}$ denote whether a user $n$ is selected as sensor in slot $t$, and $\bxt \eq \{\xnt, \forall n\in\N\} $ denote the sensor selection vector in slot $t$. 
We further denote $\bxn \eq \{\xnt, \forall t\in\T\}  $ as the allocation vector of user $n$ in all time slots. 
 


\subsection{Mobile User Modeling}
\label{sec:model-mu}

 \subsubsection{Sensing Region} 
 
\rev{Each mobile user has a certain sensing range in each time slot, mainly depending on his location and mobility pattern.} 
In Fig.~\ref{fig:model}, the sensing region of each user is illustrated by the shadow area around the user.  
Let $\znit \in \{0,1\}$ denote whether a grid $\ARi$ is located in the sensing range of  user $n$ in slot $t$. 
Then, the total \emph{sensing region} of user $n$ in each slot $t$ can be defined as a vector: 
$ \bznt \eq \{\znit, \forall i\in \I \} $. 
\rev{Note that when user $n$ moves out of the desirable sensing area in time slot $t$, we can simply define: $\znit = 0, \forall i\in \I $.} 
As mobile users move randomly, 
the sensing region $\bznt$ of each user $n$ also changes randomly across time slots.~~~~~~~~~~~~

 \subsubsection{Sensing Value} 
When a user $n$ is selected as sensor in a slot $t$, i.e., $\xnt=1$, he performs the following sensing task: collect, process, and transmit all of the data within his sensing region $\bznt$ to the SP.  
This generates a \emph{sensing value} $\vnt$ equal to the sum of weights of all grids within $\bznt$:
\begin{equation}\label{eq:vnt}
\txsty
\vnt \eq \xnt  \cdot \sumI   \znit \cdot  \wit .
\end{equation} 
 
\subsubsection{Sensing Cost}
 When performing sensing tasks, users incur extra operational cost (called \emph{sensing cost}) due to, for example,  the energy expenditure  and the transmission expense. 
Let $\cnt$ denote the total sensing cost of user $n$ in slot $t$ 
(including all potential expense used for collecting, processing, and transmitting the data within $\bznt$ to the SP). 
Obviously, such a sensing cost is user- and time-dependent. 

Due to this direct sensing cost, users may be reluctant to perform sensing tasks without sufficient incentives. 
To avoid this, in each time slot, the SP will pay certain monetary or non-monetary compensation (which we call the \emph{short-term sensing incentive})  to those users who are selected as sensors.  
Later we will show that this type of incentive can be easily addressed through, for example,   a first-degree price discrimination \cite{pricing-theory} or a truthful {auction} \cite{auction-theory} in each time slot. 


\subsubsection{{Participatory Constraint}}
As discussed in Section \ref{sec:introduction},  users may suffer certain \emph{indirect} cost even when not performing sensing tasks, induced by, for example, reporting location~/ mobility information or running sensing apps. 
Thus, if a user is rarely selected as a sensor (hence hardly receives the  short-term sensing incentive), he may gradually lose the interest in continuous participation, and decide to no longer participate in the  system (in this case, we say the user \emph{drops}). 

As shown in  \cite{Lee-2010} and \cite{Lee-2010-b}, such a long-term participation incentive strongly depends on the user's \emph{Return on Investment} (ROI). In this work, instead of directly estimating the total return and total investment, we use a simple yet representative indicator to reflect the user ROI: \emph{Allocation Probability},\footnote{Consider, for example, a user with an expected direct sensing cost $c_1$, an expected indirect sensing cost $c_2$, and an expected return $r$ when being selected as a sensor. Then, an allocation probability $\eta$ directly corresponds to an expected ROI: $\frac{r \cdot \eta}{\eta \cdot (c_1+c_2) + (1-\eta) \cdot c_2}$.} i.e., the probability of each user being selected as sensor. 
Namely, we consider such a scenario where each user $n$ will drop out of the sensing system, if his allocation probability (of being selected as sensor) is smaller than a specific threshold $D_n$, called the \emph{dropping threshold} of user $n$. 
Therefore, to ensure the active participation of users, the allocation probability of each user should be no smaller than his dropping threshold, which we call the user \emph{participatory constraint}: 
\begin{equation}\label{eq:part}
\txsty
\D_n \leq \d_n(\bxn) \eq \frac{1}{T} \sumT   \xnt , \quad  \forall n\in\N, 
\end{equation}
where $\d_n (\bxn)$ is the \revv{time average} allocation probability of user $n$, depending on the allocations of user $n$ in all slots.~~~~~~~~~~


\subsection{Service Provider Modeling}
\label{sec:model-sp}

Given the set $\N$ of mobile users participating in the system, the SP selects a subset of users as sensors in each time slot. 
We consider a \emph{non-commercial} SP (e.g., a non-profit organization or a governmental department), whose primary goal is to maximize the total sensing value and minimize the total sensing cost in the entire time period, subjecting to the user participatory constraint in (\ref{eq:part}).


Given the allocation vector $\bxt \eq \{\xnt, \forall n\in\N\} $ in slot $t$, the total sensing cost (in slot $t$) can be directly defined as the sum of all selected users' sensing costs, i.e., 
\begin{equation}\label{eq:Ct}
\txsty
\Ct \eq \sumN   \xnt \cdot  \cnt .
\end{equation}
The total sensing value (in slot $t$), however, may \emph{not}~be~same as the aggregate sensing value of all selected users due to the overlap of their sensing regions. 
The key reason is that the same data collected by multiple users simultaneously can only generate value once.  
For convenience, let $\yit$ denote whether a grid $\ARi$ is sensed by at least one mobile user, that is,~~~~~~~~~~~ 
\begin{equation}\label{eq:yit}
\txsty
\yit \eq \left \lceil \sumN   \xnt \cdot  \znit \right \rceil^{\mathbf{1}}, 
\end{equation}
where 
$\lceil x \rceil^{\mathbf{1}} = 1$ if $x \geq 1$, and $\lceil x \rceil^{\mathbf{1}} = x$ if $x < 1$. Then, the total sensing value (in slot $t$) can be defined as follows: 
\begin{equation}\label{eq:Vt}
\txsty
\Vt \eq \sumI  \yit \cdot  \wit .
\end{equation}
Obviously, if the sensing regions of all selected users do not  overlap with each other,  then $\yit = \sumN   \xnt \cdot  \znit$, and $\Vt = \sumN \vnt \cdot \xnt$, i.e., the total sensing value is directly the sum of all selected users' sensing values.~~~~~~~~~~~~~~~~~~~~

The \emph{social welfare} generated in each slot $t$ is the difference~between the total sensing value and sensing cost, i.e.,~~~~~~~~
  \begin{equation}\label{eq:VCtt}
 \VCt \eq \Vt - \Ct.
  \end{equation} 
The overall (average) social welfare in all time slots is 
  \begin{equation}\label{eq:VCt}
  \begin{aligned}
\txsty
  &  \VC(\bX) \txsty   \eq  \frac{1}{T} \sumT \VCt = \frac{1}{T} \sumT 
  \left(\Vt- \Ct \right),
  \end{aligned}
\end{equation}
where $\bX \eq  \{\xnt, \forall n\in\N, t\in\T\} \eq \{\bxt, \forall t\in\T\}$.
%
%
%

\subsection{Information Scenario}\label{sec:model:info}

We will study the sensor selection problem in different network information scenarios.
The network information consists of the weight (data value) of each grid, the sensing region and sensing cost of each user in each time slot. Formally, we define the \emph{network information} in time slot $t$ as:~~~~ 
\begin{equation}\label{eq:info}
\tht \eq \boldsymbol{\{}\wit, \bznt, \cnt,\ \forall i\in\I, n\in\N\boldsymbol{\}}.
\end{equation}
Note that the sensing value $\vnt$ of each user is not network information, as it is determined by $\wit$ and $\bznt$. 

Regarding the future network information, we consider the scenarios of \emph{complete} future information, \emph{stochastic} future information, and \emph{no} future information, depending on whether and how much the SP knows regarding the {future} network information. 
%
Regarding the current network information (realization), we consider the scenarios of information \emph{symmetry} and \emph{asymmetry}, depending on whether the SP can observe the private information of users (e.g., the sensing cost).  



\section{Off-line Sensor Selection Benchmark}\label{sec:solution}

In this section, we study the sensor selection problem with complete future information  and stochastic future information (as benchmarks). 
Note that in these benchmark cases, we assume the scenario of \emph{information symmetry} (regarding the current network information), where the SP is able to observe all of the network information realized in each time slot. 

\subsection{Complete Future Information}\label{sec:complete}

With complete future information, the SP is able to determine the sensor selections in all time slots jointly to maximize the overall social welfare. Thus, the SP's problem is~~~~~~~~~~
\begin{equation}\label{problem:complete}
\begin{aligned}
\max_{ \bX } &
 \quad \frac{1}{T} \sumT  \big( \Vt - \Ct \big) 
\\
\mbox{s.t.} & \txsty \quad \mbox{(a) } \xnt \in \{0,1\},\quad \forall n\in\N,\forall t\in\T,
\\
& 
\txsty \quad \mbox{(b) }  
\D_n \leq \d_n (\bxn) 
, \quad \forall n\in\N.
\end{aligned}
\end{equation}
\rev{Obviously, (\ref{problem:complete}) is an off-line allocation problem, and the solution presents the \emph{explicit} allocation of each user in each time slot in advance.} 
\revv{Note that (\ref{problem:complete}) is a binary integer programming, and can be effectively solved by many classic methods, such as the branch-and-bound algorithm in	\cite{report}.}~~~~~~


It is easy to see that formulating and solving (\ref{problem:complete}) requires the complete future information, which is obviously impractical. 
Hence, we will study another benchmark solution based on the stochastic information in the next subsection. 

\subsection{Stochastic Future Information}\label{sec:stochastic}

With stochastic information only, the SP cannot decide the explicit allocation of each user in each time slot in advance, due to the lack of complete future information. 
Hence, in this case, we will focus on the \emph{expected} social welfare maximization based on the stochastic information. 
 
Let $ \xn(\th) \in \{0,1\}$ denote whether a user $n$ is selected as sensor under a particular information realization $\th$,   $\bx(\th) \eq \{\xn(\th),\forall n\in\N\} $ denote the allocation vector of all users under $\th$, and $\bxn \eq \{\xn(\th), \forall \th\in\TH \} $ denote the allocation of user $n$ under all possible $\th$.
Then, the expected social welfare maximization problem can be defined as follows:\footnote{$\TH$ is the feasible set of $\th$, i.e., the set of all possible network information realizations, and $f(\th)$ is the probability distribution function (pdf) of $\th$.} 
\begin{equation}\label{problem:stochastic}
\begin{aligned}
& \max_{ \bX }  \quad \intTH  \big( \V(\th) - \C(\th) \big) \cdot f(\th) \dd \th
\\
 & \mbox{s.t.} \txsty \quad \mbox{(a) } \xn(\th) \in \{0,1\},\quad \forall n\in\N, \forall \th\in\TH,
\\
& 
\txsty \quad ~~~ \mbox{(b) }  
\D_n \leq \d_n (\bxn) , \quad \forall n\in\N, 
\end{aligned}
\end{equation}
where 
\\
$~~~~\bullet$ $\C(\th)  = \sumN   \xn(\th) \cdot  \cn(\th)$ is the sensing cost under $\th$;
\\
$~~~~\bullet $ $\V (\th) = \sumI  \yi(\th) \cdot  \wi(\th)$ is the sensing value under $\th$;
\\
$~~~~\bullet $
$\yi(\th) = \lceil  \sumN   \xn(\th) \cdot  \zni(\th) \rceil^{\mathbf{1}} $ indicates whether a grid $\ARi$ is sensed by at least one user under $\th$; 
\\
$~~~~\bullet $ $\d_n (\bxn) = \intTH  \xn(\th) \cdot f(\th) \dd \th$ is the average allocation probability of user $n$.

\rev{Similarly, (\ref{problem:stochastic}) is an off-line problem, and the solution defines a contingency plan that specifies the allocation of each user under each possible information realization $\th$ in advance.}
Note that (\ref{problem:stochastic}) is an integer programming with an infinite number of decision variables (as $\th$ is continuous), which is non-convex and NP-hard. 
Nevertheless, by the linear programming relaxation, we can easily transform (\ref{problem:stochastic}) into a linear programming problem, and solve it by classic methods, e.g., the KKT analysis.\footnote{We leave the details in \cite{report}, as the method is standard. Moreover, solving the stochastic opitmalization problem is not the main contribution of this work.}~~~~~~~~

Next we analyze the gap between the maximum social welfare with complete information (denoted by $\VCo$) derived from (\ref{problem:complete}) and the maximum expected social welfare with stochastic information (denoted by $\VCso$) derived from (\ref{problem:stochastic}).~Formally, 

\begin{lemma}
\label{lemma:gap}
If $T\rightarrow \infty$, then
$
\VCso \rightarrow \VCo.
$
\end{lemma}


Lemma \ref{lemma:gap} indicates that as long as the total sensing period $T$ is large enough, the social welfare loss induced by the loss of complete network information is negligible.
Hence, both $\VCo $ and $\VCso $ will serve as the same benchmark for the on-line policies proposed in Sections \ref{sec:noinform} and \ref{sec:asymmetric}.  

It is notable that formulating and solving (\ref{problem:stochastic}) still requires certain (stochastic) future information, which may not be available in practice. 
This motivates us to further study on-line policies not relying on any future  information.~~~~~~~~~~~~~~~~~~~


\section{On-line Sensor Selection Policy}\label{sec:noinform}



In this section, we study the sensor selection problem with no future information. 
We propose an on-line sensor selection policy based on the \emph{Lyapunov optimization} framework \cite{Neely}, which relies only on the current network information and past sensor selection history, while not on any future information. 
Meanwhile, it asymptotically converges to the benchmark with complete or stochastic future information proposed in Section \ref{sec:solution}.
\rev{Note that we also assume the scenario of information symmetry in this section, and will further study the scenario of information asymmetry in the next section.} 



\subsection{{Lyapunov Optimization Technique}}


Lyapunov optimization \cite{Neely} is a widely used technique for solving stochastic optimization problems with time average constraints, such as the   social welfare maximization problem (\ref{problem:complete}) in this work (with $T\rightarrow \infty$), where the user participatory constraint (b) is the time average constraint.
Hence, we introduce the Lyapunov optimization technique to solve the sensor selection problem (\ref{problem:complete}) with no future information.

\subsubsection{Queue Definition}

The key idea of Lyapunov optimization technique is to use the \emph{stability} of queues to ensure that the time average constraints are satisfied. 
Following this idea, we first introduce a \emph{virtual queue} ($\quen$) for each   user $n$. This virtual queue is used for buffering the \emph{virtual}  allocation request of each user.
Here, we use the prefix ``virtual'' to denote that the request is not actually initiated by the user, but rather, it is used to reflect the requirement of the user participatory constraint. 
Namely, one virtual request represents that ``\emph{to satisfy the user participatory constraint, the~user~\textbf{should}~be~selected~as~sensor in one additional~time~slot}''. 
Hence, the backlog of a virtual queue denotes the total number of virtual requests in the queue (which may not be an integer), i.e., the total number of additional time slots that the user should be selected as sensor
(in order to meet his participatory constraint).~~~~~~~~~~~~~~~~~~~~~~~~~~~~~~~~~

\subsubsection{Queue Dynamics}

With the above queue definition, each virtual request of user $n$ will enter into the queue with a constant \emph{arrival rate} of $\D_n$.
Let $\xnlo [t] \in\{0, 1\}$ denotes whether user $n$ is selected as sensor in time slot $t$ (under certain sensor selection policy), and $\dnlo \eq \dn(\bxnlo) = \frac{1}{T} \sumT \xnlo[t] $ denote the average allocation probability of user $n$. 
Intuitively, $\xnlo [t] =1 $ implies that one virtual request of user $n$ leaves the queue at slot $t$. 
Hence, 
the virtual request of user $n$ will leave the queue with an average \emph{departure rate} of $\dnlo$.   

Let $\quen^t$ denote the queue backlog of user $n$ in slot $t$, 
and let $\bque^t \eq \{\quen^t,\forall n\in\N\}$ denote the queue backlog vector of all users. 
For each user $n$, given the constant arrival of his virtual request and the allocation $\xnlo[t]$ in each slot $t$ (departure), we have the following dynamic equation for his virtual queue:
\begin{equation}\label{eq:queue-dynamics}
\quen^{t+1} = \left[\quen^{t} - \xnlo [t]  \right]^+  + \D_n, 
\end{equation}
where $[x]^+ = \max(x, 0)$.


Next, we show how to connect the queue stability condition with the user participatory constraint in our problem.
We say a virtual queue $\quen$ is \emph{rate stable}, if 
$$
\lim_{t\rightarrow \infty} \frac{\quen^t }{t} = 0 \mbox{~~with probability 1}.
$$
By the queue stability theorem \cite{Neely}, a queue $\quen$ is rate stable if and only if the arrival rate is no larger than the departure rate, i.e., $\D_n\leq \dnlo$. 
\emph{This establishes the equivalence between the queue stability condition and the user participatory constraint.}
That is, to guarantee the user participatory constraint in our problem, we only need to ensure that the associated virtual queue is rate stable under the proposed policy.

\subsubsection{Queue Stability}

Now we study the queue stability 
 using the Lyapunov drift.
We first define the Lyapunov function:
\begin{equation}\label{eq:lyapunov}
\lyp[t] \eq \frac{1}{2} \sumN (\quen^t)^2. 
\end{equation}
The \emph{Lyapunov drift} in each slot $t$ is defined as the change of Lyapunov function from one slot to the next, i.e.,~~~~~ 
\begin{equation}\label{eq:drift}
\drift[t] \eq \lyp[t+1] - \lyp[t]. 
\end{equation}
By the Lyapunov drift theorem (Th.~4.1 in \cite{Neely}), if a policy \emph{greedily minimizes the Lyapunov drift $\drift[t]$} in each slot $t$, then all backlogs are consistently pushed towards a low level, which potentially maintains the stabilities of all queues (i.e., ensures the participatory constraints of all users). 

\subsubsection{Joint Queue Stability and Welfare Maximization}

Next, we analyze  the joint queue stability and objective optimization (i.e., expected social welfare maximization). 
By the Lyapunov optimization theorem (Th.~4.2 in \cite{Neely}), to stabilize the queues while optimizing the objective, we can use such an allocation policy that greedily minimizes the following \emph{drift-plus-penalty}: 
\begin{equation}\label{eq:drift-penalty}
\pnlt[t] \eq 
\drift[t] - \lypV \cdot 
\big(\Vt - \Ct\big), 
\end{equation}
where the (negative) social welfare, i.e., $ \Ct - \Vt$, is viewed as the penalty incurred on each slot $t$;   
$ \lypV \geq 0$ is a non-negative control parameter that is chosen to achieve a desirable tradeoff between the optimality and queue backlog.

We further notice that directly minimizing the drift-plus-penalty defined in (\ref{eq:drift-penalty}) may be difficult (partly because $\drift[t] $ is a quadratic function). Hence, we will focus on minimizing a specific \textbf{upper-bound} of  the drift-plus-penalty to achieve the joint stability and optimization.  

Next, we give such an upper-bound. Notice that
\begin{equation} 
\begin{aligned}
\drift[t] & \txsty   \leq  \frac{1}{2} \sumN \left( 
\xnlo[t]^2 + \D_n^2 + 2 \cdot \quen^{t} \cdot (\D_n - \xnlo[t])
 \right)
 \\
 &  \txsty   \leq B  + \sumN \quen^{t} \cdot (\D_n - \xnlo[t]),
\end{aligned}
\end{equation}
where $B \eq \sumN \frac{1 + \D_n^2 }{2}$ is a constant.\footnote{The first inequality follows because $([q - x]^+ + D)^2 \leq q^2 + x^2 + D^2 + 2 q\cdot (D-x) $. 
The second inequality follows because $\xnlo[t]^2 \leq 1$.} 
Then, we have the following upper-bound for the drift-plus-penalty in (\ref{eq:drift-penalty}): 
\begin{equation}\label{eq:penalty-bound}
\pnlt[t] 
\leq B + \sumN \quen^{t} \cdot (\D_n -   \xnlo[t]) - \lypV \cdot 
\big(\Vt - \Ct\big). 
\end{equation}
By the Lyapunov optimization theory, it is easy to show that minimizing the above {upper-bound} of the drift-plus-penalty is equivalent to minimizing the   drift-plus-penalty itself, in terms of the queue stability and objective optimization.

\rev{
\emph{Remark.} Beyond following the standard Lyapunov optimization framework \cite{Neely}, our own contributions in this part are two-fold. First, we explicitly define the virtual queue, and analytically connect the user participatory constraint and the queue stability. This is the basis of applying Lyapunov optimization in our problem. 
Second, we propose an upper-bound (\ref{eq:penalty-bound}) for the drift-plus-penalty, which is problem-specific and does not have a generic form suitable for all problems. 
The later on-line policy is based on this upper-bound.}


\begin{algorithm}[t]
\label{algo:2}
\small
\caption{Lyapunov-based Policy (Information Symmetry)}
\DontPrintSemicolon
\textbf{Initialization:} $\bque = \bque^0$;\;
\For{each time slot $t=0,1,...,T$}
{
\emph{Allocation Rule:}
\begin{equation*}
\begin{aligned}
\bxlo[t]
& = \arg \max_{ \bxt } \Big(
\Vt - \Ct + \txsty  \sumN \frac{\quen^{t}}{\lypV} \cdot \xnt \Big)
\end{aligned}
\end{equation*}
\emph{Updating Rule:}
$$
\txsty
\quen^{t+1}= \left[\que^{t} - \xnlo[t] \right]^+ + \D_n, \quad \forall n\in\N 
$$
}
\end{algorithm}

\subsection{{On-line Allocation Policy}}

Based on the above theoretical analysis, we now design an on-line policy that aims at  minimizing the  drift-plus-penalty upper-bound in (\ref{eq:penalty-bound}) in each time slot. 
We present such a Lyapunov optimization based policy in Policy \ref{algo:2}.~~~~~~~~~~~~

\subsubsection{Algorithm Design}

The proposed Policy \ref{algo:2} consists of an \emph{allocation rule} and an \emph{updating rule} in each time slot. 
The allocation rule determines the sensor selection (allocation) $\bxlo[t]$ in each slot $t$, based on the current network information $\th[t]$ and the current queue backlogs $\bque^t$, aiming at minimizing the upper-bound of drift-plus-penalty in (\ref{eq:penalty-bound}).
The updating rule updates the queue backlogs based on the current allocation result $\bxlo[t]$ according to (\ref{eq:queue-dynamics}).   
It is easy to see that Policy \ref{algo:2} relies only on the current network information and the past sensor selection history (captured by the queue backlogs), while not on any complete or stochastic future information.

%
%

%


\subsubsection{Optimality}

Now we provide the optimality of Policy \ref{algo:2}. 
Let $\VClo[t] $ denote the social welfare generated in each slot $t$, and $\VCso$ denote the maximum social welfare benchmark with the stochastic information (derived in Section \ref{sec:solution}).  Formally,

\begin{theorem}[Optimality]\label{theorem:2} 
$$
 \lim_{T\rightarrow \infty} \frac{1}{T} \sumT \mathbf{E} (\VClo[t])  \geq  \VCso  -  \frac{B}{\lypV}.
$$
\end{theorem}

The proof follows standard Lyapunov optimization theory \cite{Neely}. 
By Theorem \ref{theorem:2}, we can easily find that Policy \ref{algo:2} converges to the maximum social welfare benchmark asymptotically, with a \emph{controllable} approximation error bound $O(1/{\lypV})$. 

Intuitively, in Policy \ref{algo:2}, each virtual queue can be viewed as a \emph{regulation factor} for lowering (regulating) the sensing cost of that user, and hence increasing the selection probability of that user. 
By the updating rule in Policy \ref{algo:2}, we can further obtain \revvv{the following approximation for the queue backlog}\footnote{\revvv{This approximation} is obtained by simply omitting the operation $[.]^+$.}  
$$
\quen^{t} \approx \quen^0 - \sum_{k=0}^{t-1} \xnlo[k] + t \cdot D_n. 
$$
This implies that the time-attenuated queue backlog $\frac{\quen^{t}}{t}$ can be used to approximate the gap between the required allocation probability (i.e., $D_n$) and the actual allocation probability till slot $t$ till slot $t$  (i.e., $\frac{\sum_{k=0}^{t-1} \xnlo[k]}{t}$). 
\rev{Notice that the queue backlog $ {\quen^{t}} $ is bounded, hence the above gap goes to zero as $t \rightarrow \infty$.} 




\section{Auction-based On-line Sensor Selection Policy}\label{sec:asymmetric}


In this section, we consider the asymmetric information scenario (regarding the current information), 
where \emph{the sensing cost of each user $n$ realized in each time slot $t$ (i.e., $ \cnt $) is his private information}, and cannot be observed by the SP. 
Obviously, without this private sensing cost, the SP cannot implement the allocation rule in Policy \ref{algo:2}.

\subsection{Auction Mechanism Design}

We design an (reverse)  VCG auction  to address the credible information disclosure of users in each time slot, 
where the SP is the auctioneer (buyer), and users are the bidders (sellers). 
A standard VCG auction usually consists of an \emph{allocation rule} (winner determination) and a \emph{payment rule}. 
Our proposed auction mechanism involves a set of regulation factors (which are introduced for ensuring the user participatory constraint), hence includes an additional \emph{updating rule} for the regulation factors. 
We present the detailed auction mechanism in Policy \ref{algo:3}. 
Next we will explain these rules in details. 

For convenience, we denote $\bidnt $ as each user $n$'s \emph{bid} (report) regarding his sensing cost in each slot $t$, and $\regun^t$ as the \emph{regulation factor} (similar as the virtual queue in Section \ref{sec:noinform}) associated with each user $n$ in each slot $t$. 

\subsubsection{\textbf{Allocation Rule}}

The allocation rule aims at maximizing a \emph{regulated} social welfare in each time slot: 
$$
\WVC[t] \eq \Vt - \sumN  \xnt \cdot \wcnt,  
$$
where $\wcnt \eq \bidnt - \regun^{t}$ is the regulated sensing cost of user $n$, depending on both the user bid and the regulator factor. 
For convenience, we denote $\bxau[t] \eq \{\xnau[t], \forall n\in\N\} $ as the allocation result in slot $t$ (i.e., that maximizes $\WVC[t]$). 

%

\begin{algorithm}[t]
\label{algo:3}
\small
\caption{Auction-based Policy  (Information Asymmetry)}
\DontPrintSemicolon
\textbf{Initialization:} $\bregu = \bregu^0$;\;
\For{each time slot $t=0,1,...,T$}
{
Denote   $\bidnt$ as the bid of each user $n$;\;
\emph{Allocation Rule:}
\begin{equation*}
\begin{aligned}
\bxau[t]
& = \arg \max_{ \bxt } \ 
\Vt - \sumN  \xnt \cdot (\bidnt - \regun^{t})  \end{aligned}
\end{equation*}
\emph{Payment Rule:}
\begin{equation*}
\pnt =  \xnau[t] \cdot \Big( \V^\ddag[t] - \C^\ddag_{-n}[t] -\WVC^{\sharp}_{-n}[t] + \regun^{t} \Big)
\end{equation*}
\emph{Updating Rule:}
\begin{equation*}
\begin{aligned}
\txsty
\textstyle
\regun^{t+1}= \frac{1}{\lypV} \cdot 
\left(\left[\lypV\cdot\regun^{t} - \xnau[t] \right]^+ + \D_n\right),   ~\forall n\in\N 
\end{aligned}
\end{equation*}
}
\end{algorithm}

\subsubsection{\textbf{Payment Rule}}

The payment to user $n$ in each time slot $t$ is: (i) $\pnt = 0$ if user $n$ is not selected, i.e., $\xnau[t]=0$, or (ii) if user $n $ is selected, i.e., $\xnau[t]=1$, then
\begin{equation}\label{eq:auction-pay}
\pnt =   \V^{\ddag}[t] - \C^{\ddag}_{-n}[t] -\WVC^{\sharp}_{-n}[t] + \regun^{t}, 
\end{equation}
where $\V^{\ddag}[t]$ is the total sensing value under $\bxau[t]$, $\C^{\ddag}_{-n}[t] = \sum_{k\neq n} \x^{\ddag}_k[t] \cdot \widetilde{c}_k[t] $ is the total sensing cost except that of user $n$ under $\bxau[t]$, and  
$\WVC^{\sharp}_{-n}[t] $ is the maximum achievable social welfare when excluding user $n$ in the system. 
The first 3 terms correspond to the payment in a standard VCG auction. 
The last term is used to compensate the user cost regulation.~~~~~

\subsubsection{\textbf{Updating Rule}}
 
Inspired by Policy \ref{algo:2}, we have:
$$
\textstyle
\regun^{t+1}= \frac{1}{\lypV} \cdot \left(\left[\lypV\cdot\regun^{t} - \xnau[t] \right]^+ + \D_n\right),   ~\forall n\in\N.
$$
The above updating rule is exactly same as that in Policy \ref{algo:2}, by simply viewing $\lypV\cdot\regun^{t} $ as $\quen^t$. 
Obviously, if users are truthful, then the above allocation/updating rule achieves the exactly same allocation and performance as in Policy \ref{algo:2}.

\subsection{Truthfulness and Optimality}
 


\begin{theorem}[Truthfulness]\label{lemma:truthfulness}
The auction in Policy \ref{algo:3} is truthful.  
\end{theorem}

\begin{proof}
Due to the space limit, we only show that each user $n$ has no incentive to report (bid) a cost \emph{higher than} his true cost.\footnote{The proof for ``users are not willing to report costs lower than the true values'' is similar. Please refer to \cite{report} for details.} 
There are 4 possible outcomes:

\noindent
(a) \{\emph{loss, loss}\}: user $n$ loses when bidding both truthfully and non-truthfully. He receives a zero payment in both strategies.~~~~ 

\noindent
(b) \{\emph{win, loss}\}: user $n$ wins (loses) when bidding truthfully (non-truthfully). He receives a smaller payment (i.e., zero) when bidding non-truthfully.
 
\noindent 
(c) \{\emph{loss, win}\}: user $n$ loses (wins) when bidding truthfully (non-truthfully). This is practically impossible, as a user losing with a lower cost will never win when submitting a higher cost. 

\noindent
(d) \{\emph{win, win}\}: user $n$ wins when bidding both truthfully and non-truthfully. We will show that user $n$ receives the same payment in both strategies.  
First, the third term and the last term in (\ref{eq:auction-pay}) are obviously identical in both strategies. 
Second, the first two terms in (\ref{eq:auction-pay}) are also identical due to the following assert: \emph{If an allocation vector $\bx^*$ maximizes the social welfare, then, excluding any user $n$ and removing the grids sensed by user $n$ (under $\bx^*$), the remaining vector $\bx_{-n}^*$ maximizes the social welfare in the remaining system.} 
\end{proof}

\begin{theorem}[Optimality]
The auction in Policy \ref{algo:3} achieves the same asymptotically optimal social welfare as in Policy \ref{algo:2}.
\end{theorem}

\begin{proof}
By the truthfulness given in Theorem \ref{lemma:truthfulness}, together with the observation that the allocation and updating rules in Policy \ref{algo:3} are exactly same as those in Policy \ref{algo:2}, we can prove the optimality immediately. 
\end{proof}



%
%
%
 
\rev{
\emph{Remark.}
The above Policy \ref{algo:3} is truthful only when users are \emph{myopic}, in the sense that they only care about the current benefits in each time slot, while not anticipating the potential impacts of their bidding strategies on the future benefits. 
As a counter-example, a non-myopic user may report a large fake cost.  
By doing so, the SP will assign a large regulation factor to the user (in order to satisfy the user's participatory constraint), which potentially increases the user's future payment. 
We will study the model with non-myopic users in our future work.
}


\section{Simulations}\label{sec:simulation}

In our simulations, we launch a participatory sensing application in a middle-scale virtual city with size $10\mathrm{km}\times10\mathrm{km}$. 
The whole area is divided into $2500$ grids, each corresponding to a square of $200\mathrm{m}\times 200\mathrm{m}$. 
\rev{Users move according to the random walk model: in each time slot, each user jumps from the original location (grid) to another location (grid) randomly according to certain probability distribution.  
For illustrative purposes, we assume that 
the sensing region of each user in each time slot is a disk, centered at his location, with a radius randomly picked from $[400\mathrm{m}, 800\mathrm{m}]$.}\footnote{We will consider the more practical mobility model and sensing region scenario based on real data traces in our future work.}
We run the system in a period of $10,000$ time slots, \rev{which is long enough for obtaining stable outcomes under our adopted policies.}


\begin{figure}[t]
\centering
\includegraphics[scale=0.8]{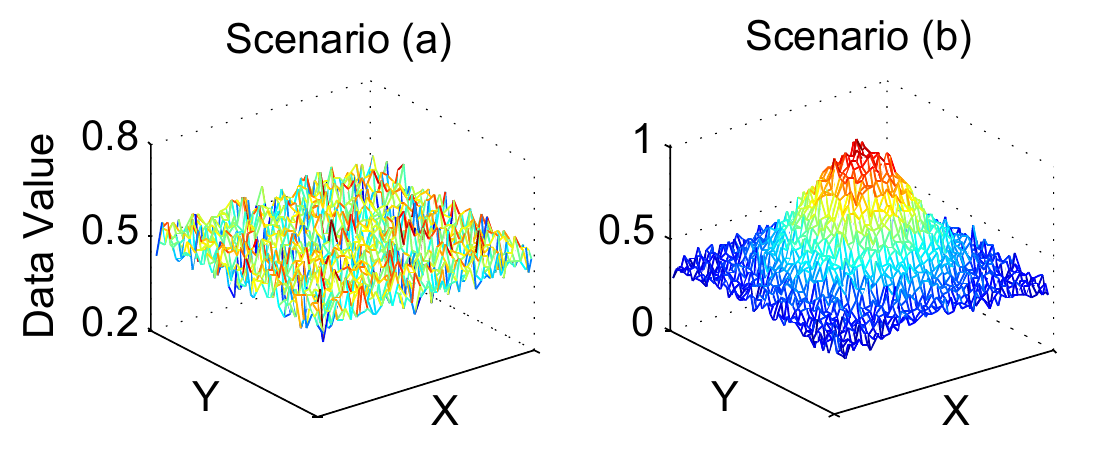}
\vspace{-3mm}
\caption{Illustration of Two Scenarios: (a) no hotspot and (b) one hotspot.}
\label{fig:scenarios}
\end{figure}

\subsection{Simulation Scenarios}

We consider two different simulation scenarios (a) and (b), depending on the different data value distributions in different areas, as shown in Fig.~\ref{fig:scenarios}. 
In scenario (a), there is no hotspot, 
and all grids are of the similar importance. 
Hence, the data value in different areas follows an {i.i.d.} distribution. 
In scenario (b), there is one hotspot, and the grids near to the centre of the hotspot are more important than those far from the hotspot, and hence have larger data values. 
\rev{Note that any scenario with multiple hotspots can be viewed as an intermediate case between (a) and (b).}  
For fair comparison, we set the average data value in the whole area as 0.5 for both scenarios. 

%
%

\begin{figure*}[t]
\centering
\includegraphics[scale=0.67]{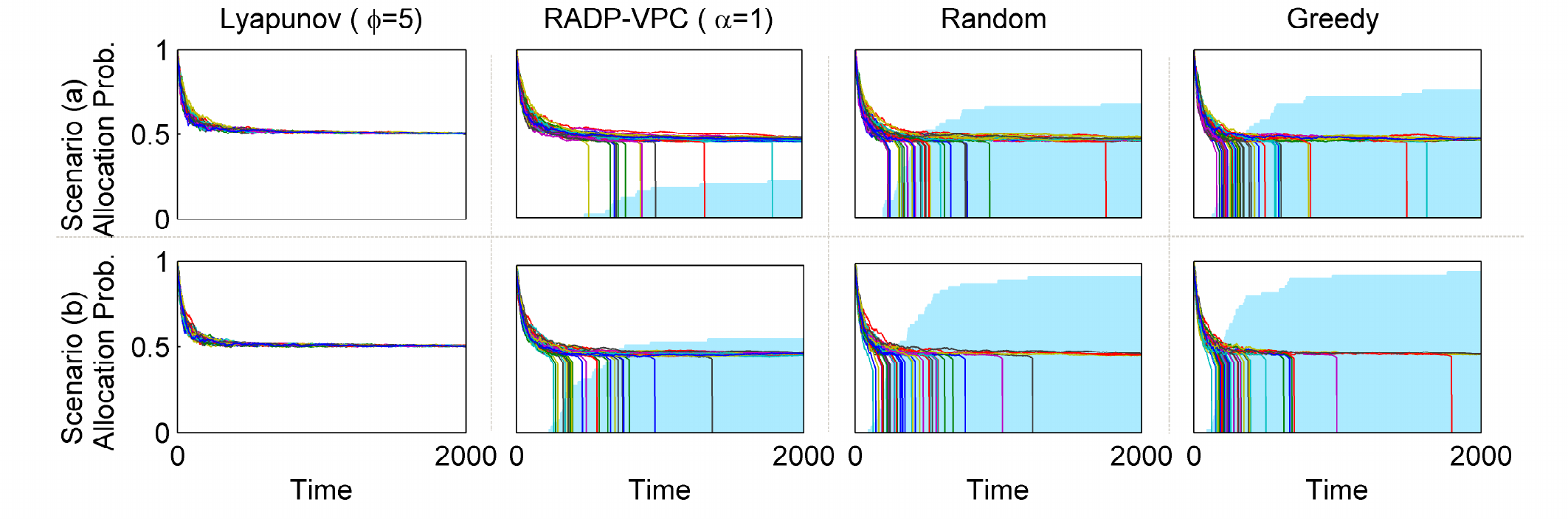} 
\vspace{-3.5mm}
\caption{Allocation Probability Dynamics and User Dropping in Scenario (a) (the first row) and Scenario (b) (the second row).
The dropping of a user is illustrated by the sudden decrease of his allocation probability. 
The percentage of dropping users is denoted by the blue shadow area.}
\label{fig:dropping}
\end{figure*}

\begin{figure*}[t]
\centering
\includegraphics[scale=0.6]{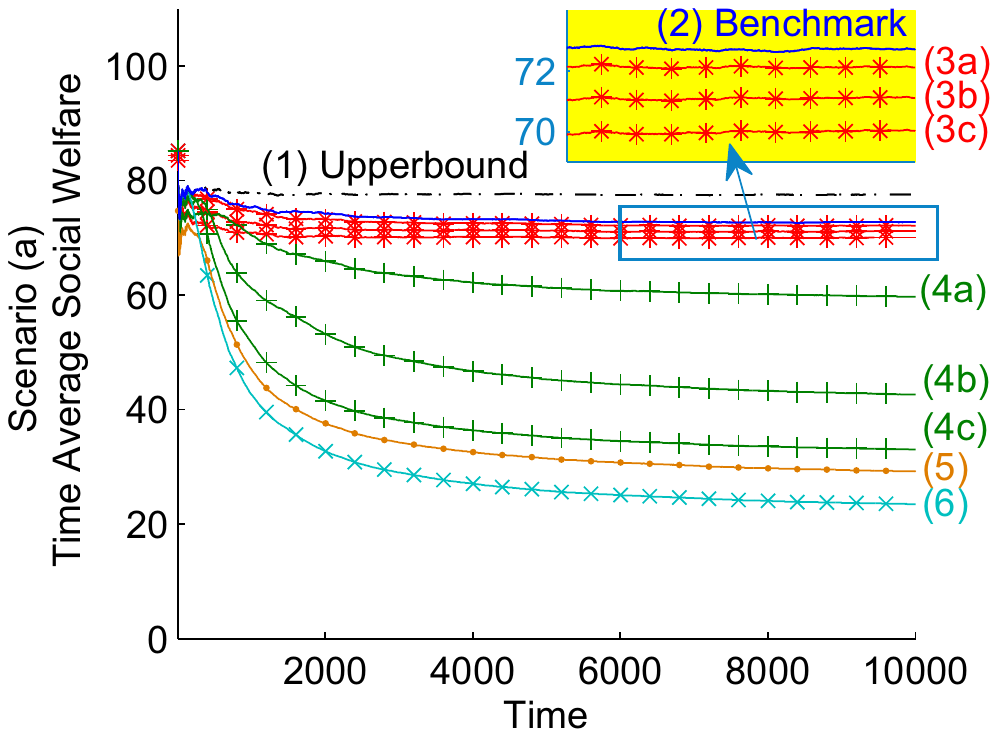}
~~~~~
\includegraphics[scale=0.6]{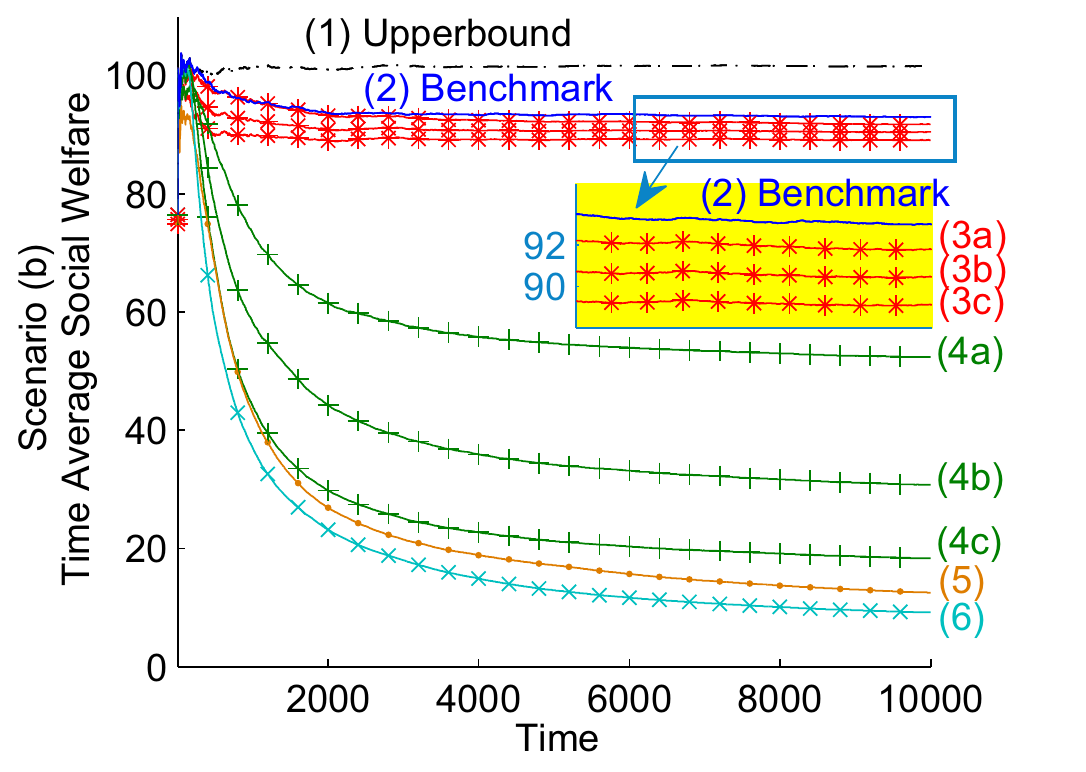}
\vspace{-2.5mm}
\caption{Average Social Welfare under Different Policies in Scenario (a) (left) and Scenario (b) (right). 
Policy: 
(3a)-(3c) \emph{Lyapunov-based policy} ($\lypV = \{20, 10, 5\}$) proposed in this work;
(4a)-(4c) \emph{RADP-VPC policy} ($\alpha = \{1, 0.5, 0.2\}$) proposed in \cite{Lee-2010} and \cite{Lee-2010-b};
(5) \emph{Random policy}; 
(6) \emph{Greedy policy}.
}
\label{fig:sw}
\end{figure*}

\subsection{Performance Comparisons}


Now we compare the performance of our proposed policy with the RADP-VPC policy proposed in \cite{Lee-2010} and \cite{Lee-2010-b}, a well-known policy that considers the participation incentive.\footnote{\revvv{In the RADP-VPC policy, each user $n$'s cost is regulated by a virtual credit $v_n$, and the virtual credit $v_n$ is updated in the following way: (i) $v_n = v_n + \alpha$ if user $n$ is not selected as sensor in the previous slot, and (ii) $v_n = 0$ if user $n$ is selected as sensor in the previous slot, where $\alpha > 0$ is a controllable parameter. Intuitively, a larger $\alpha$ can better satisfy the user participatory constraint, but may reduce the generated social welfare.}}
To draw a more convincing conclusion, we also compare our policy with those not considering the participation incentive, e.g., random selection and greedy selection  (both are widely used in practical applications such as Waze\cite{App-Waze}).\footnote{In the greedy (random) policy, users are selected one by one in a descending (random) order of their  generated social welfares.}

\subsubsection{\textbf{Dropping Probability}} 
We first compare the user dropping probability under different policies. 
In this simulation, we set the dropping threshold as 0.5 for all users. 
Namely, if the allocation probability of a user is less than 0.5, the user will drop out of the system.\footnote{\revvv{To reduce the ``start effect'' where a user may mistakenly drop in the first few slots (due to the low allocation probability in these slots), we assume that all users will be selected as sensor in the first 40 time slots.}} 
Fig.~\ref{fig:dropping} illustrates the dynamics of user allocation probabilities as well as the dropping of users.
We can see that, in scenario (a)  (the first row), more than $70\%$ of users drop  under the greedy or random sensor selection policy, and around $25\%$ of users drop   under the RADP-VPC policy ($\alpha = 1$); and in scenario (b)  (the second row), more than $90\%$ of users drop  under the greedy or random sensor selection policy, and more than  $50\%$ of users drop  under the RADP-VPC policy ($\alpha = 1$). 
Our proposed policy, however, retains all users in both scenarios. 
 
\revvv{We can further see that under the same policy (except our proposed one),  more users drop in the scenario (b) than in scenario (a). The reason is as follows. 
In scenario (b) with one hotspot, most of the data value is concentrated in the hotspot area, and hence a large total sensing value can potentially be collected by a small number of  users (located in the hotspot area).
In scenario (a) with no hotspot, however, the data value is uniformly distributed in all areas, and hence a large total sensing value can   be collected only by a large enough number of users (distributed in the whole area). 
Hence, to achieve the same level of sensing value, the number of sensors needed in scenario (b), on average, is smaller than that needed in scenario (a). 
Accordingly, the user allocation probability is lower, and hence more users drop,   in scenario (b)}.

\begin{figure*}[t]
\centering
\begin{minipage}[t]{0.325\linewidth}
\centering
\includegraphics[scale=0.3]{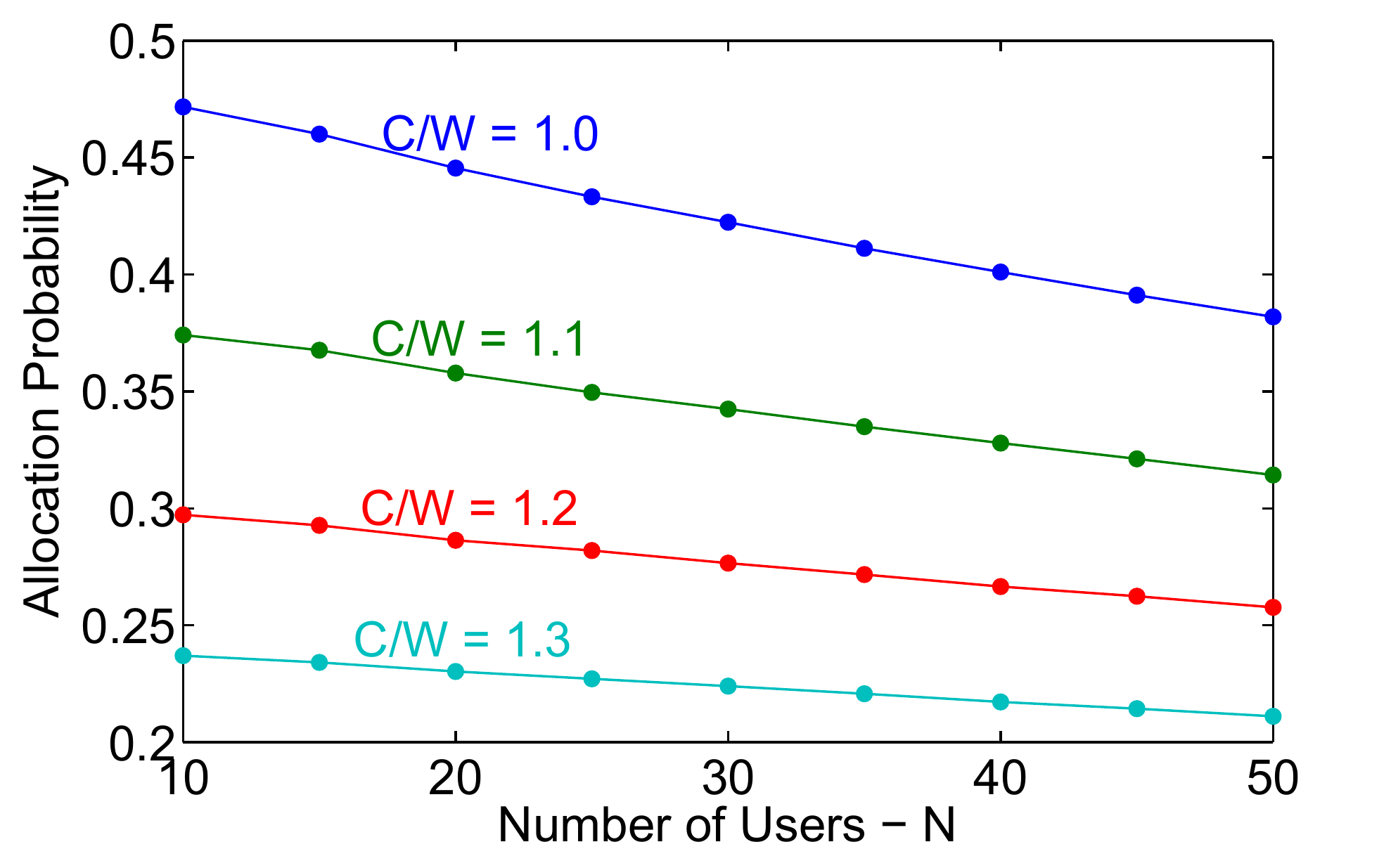}
\caption{Allocation Probability vs User Number }
\label{fig:impact1}
\end{minipage}
\begin{minipage}[t]{0.325\linewidth}
\centering
\includegraphics[scale=0.3]{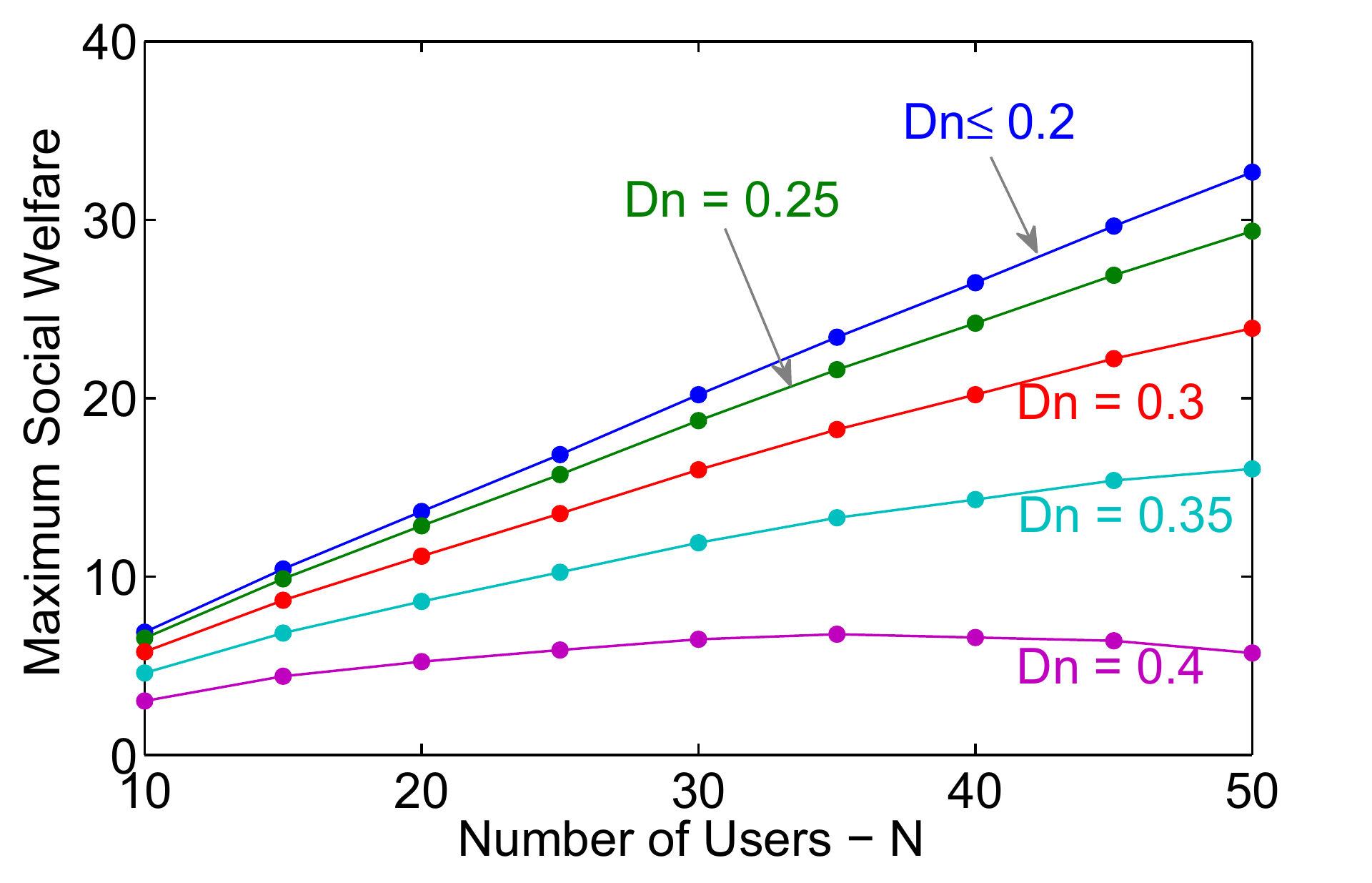}
\caption{Social Welfare vs User Number}
\label{fig:impact2}
\end{minipage}
\begin{minipage}[t]{0.327\linewidth}
\centering
\includegraphics[scale=0.3]{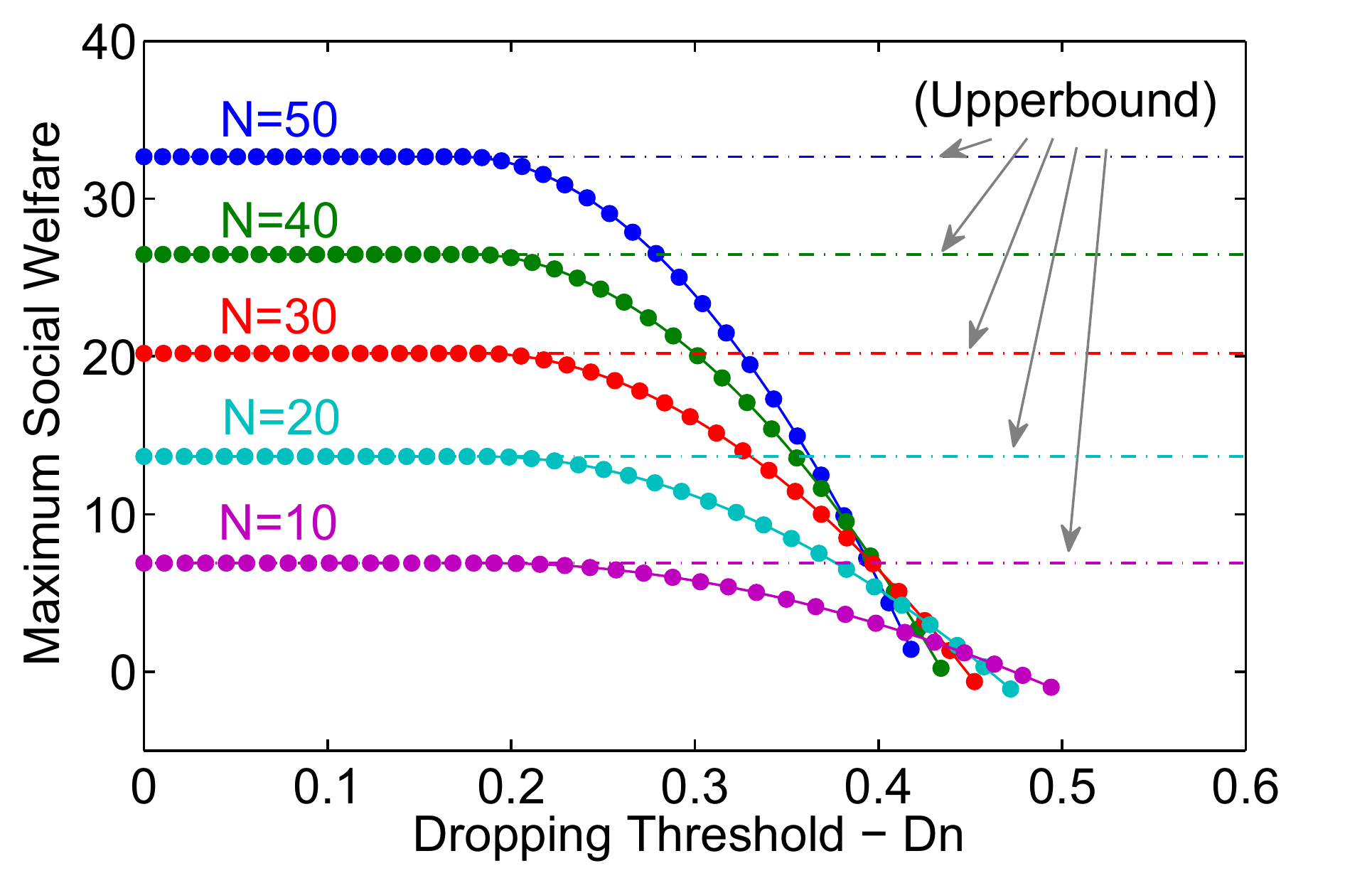}
\caption{Social Welfare vs Dropping Threshold}
\label{fig:impact3}
\end{minipage}
\end{figure*}

\subsubsection{\textbf{Social Welfare}}
We then compare the  average social welfare under different policies in Fig.~\ref{fig:sw}. 
Curve (1) is the maximum social welfare with \emph{no} participatory constraint, and serves as an upper-bound of the maximum achievable social welfare with the participatory constraint. 
Curve (2) is the maximum social welfare benchmark (with the participatory constraint) with complete or stochastic future information derived in Section \ref{sec:solution}. 
The gap between curves (1) and (2) is   called the \emph{incentive cost}, which is used to guarantee the user long-term participation incentive. 
In our simulations, the incentive cost is approximately $6\%$ in scenario (a) and $8\%$ in scenario (b). Namely, the incentive cost is higher in scenario (b) than (a) due to the higher dropping probability.~~~~~~~~~~~~~~~~~~~~~~~~~~~~~~~~ 

Curves (3a)-(3c) denote the social welfares achieved by our proposed Lyapunov-based Policy \ref{algo:2} or \ref{algo:3} (with $\lypV = 20$, $10$, and $5$, respectively).
Our policy converges to the optimal benchmark asymptotically, with very small approximation errors, e.g., $\{1\%,\ 2\%,\ 3\%\}$ in scenario (a) and $\{1.5\%,\ 3\%,\ 4.5\%\}$ in scenario (b). 
Note that the approximation error bound is controllable, via choosing different values of $\lypV$. 
We can further see that the benchmark (i.e., the maximum social welfare) is higher in scenario (b), as the same amount of sensing value can potentially be collected by fewer users in scenario (b) than in scenario (a). 
Accordingly, our proposed policy can achieve a higher social welfare in scenario (b).~~~~~~~~~~~~~~~~~~~~~~~~~~~~~~~~~

Curves (4a)-(4c) denotes the social welfares achieved by the RADP-VPC policy (with $\alpha = 1$, $0.5$, and $0.2$, respectively) proposed in \cite{Lee-2010} and \cite{Lee-2010-b}. 
Obviously, the performance of RADP-VPC largely depends on the choice of parameter $\alpha$. 
In scenario (a), 
the social welfare gap between the RADP-VPC policy and our policy ranges from $15\%$ (when $\alpha=1$) to $50\%$ (when $\alpha =0.2$). 
In scenario (b), this gap increases to $40\%$ and $75\%$. 
In fact, different from our policy or benchmark, the RADP-VPC policy achieves a worse performance in scenario (b), due to the higher dropping probability in scenario (b). 
This illustrates the importance of considering the long-term participatory incentive in a sensing system.~~~~~~~~~~~~~~~~~~~~~~~~~~~~~~~~~~~~~~~

Finally, Curves (5) and (6) denotes the social welfares achieved by the random and greedy sensor selection policies. 
Neither policy considers the long-term participation incentive, hence users drop quickly (see Fig.~\ref{fig:dropping}) and  the social welfare decreases dramatically. 
The social welfare gap between the these two policies and our policy is larger than $60\%$ in scenario (a) and $85\%$ in scenario (b). 
Similarly to the RADP-VPC policy, these two policies both achieves a worse performance in scenario (b) than in (a), due to the higher dropping probability in scenario (b). 
Counter-intuitively, the greedy policy achieves a worse performance than the random policy, due to the higher user dropping probability in the greedy policy. 
This also illustrates the importance of considering the long-term participatory incentive in a sensing system.

%

\subsection{Impact of Participatory Constraint}

So far, we have   shown in Fig.~\ref{fig:sw} that our policy converges to the maximum social welfare benchmark asymptotically. 
Next, we show in Figs.~\ref{fig:impact1}-\ref{fig:impact3} that how the participatory constraint affects this benchmark. 
We provide the results in scenario (a) only, as those in scenarios (b)-(d) are similar.~~~~~~~ 

Fig.~\ref{fig:impact1} illustrates the user allocation probability vs the number of participating users, under~different~sensing~costs~(where C/W denotes the average ratio of 
unit sensing cost and unit data value). 
Obviously, the allocation probability decreases with both the sensing cost and the number of users (due to the partial conflict of their sensing activities). 
Note that in this result, there is no participatory constraint. 
Namely, users never drop, and in each time slot they will be selected based on the realized costs. 
This result is useful for explaining the different impacts of participatory constraint discussed later.~~~~

Fig.~\ref{fig:impact2} illustrates the maximum social welfare (benchmark) vs the number of participating users $N$, under different dropping thresholds. 
We can see that when the dropping threshold is small (e.g., $D_n \leq 0.35$), the maximum social welfare always increases with the number of users, and the increase rate becomes larger with a smaller dropping threshold. 
When the dropping threshold is large (e.g., $D_n \geq 0.4$), the maximum social welfare first increases with the number of users, and then decreases with the number of users. 
This implies that in a sensing system with a mild or no participatory constraint (e.g., a small or zero dropping threshold), 
we can always increase the social welfare by involving more users into the sensing system. 
With a stringent  participatory constraint (e.g., a large dropping threshold), however, involving more users may not always increase the social welfare, due to the high incentive cost to retain users in the system. 

Fig.~\ref{fig:impact3} illustrates the maximum social welfare (benchmark) vs the dropping threshold $D_n$, with different numbers of users. 
Each dash line denotes the maximum social welfare without the participatory constraint (i.e., the upperbound in Fig.~\ref{fig:sw}). 
We can see that the social welfare decreases with the dropping threshold, as a higher dropping threshold implies that more incentive cost is needed to retain the users in the system. 
We can further see that such a welfare degradation (induced by the incentive cost) is more severe with a larger number of users, as the total incentive cost increases with the number of users. 

%
%



\section{Conclusion}\label{sec:conclusion}

In this work, we studied the optimal sensor selection problem in a general time-dependent and location-aware participatory sensing system with the user long-term participatory constraint. 
We proposed Lyapunov based on-line sensor selection (auction) policies, which do not rely on future information and achieve the optimal off-line benchmark performance asymptotically. 
There are several possible extensions in the future work. 
An interesting one is to study the truthful mechanism when users are \emph{not} myopic and can somehow anticipate the impact of their activities on the future time slots.



%
%


\end{document}